# Kinematics clustering enables head impact subtyping for better traumatic brain injury prediction.

Xianghao Zhan(1), Yiheng Li(2), Yuzhe Liu(1), Nicholas J. Cecchi(1), Olivier Gevaert(2), Michael M. Zeineh(3), Gerald A. Grant(4), David B. Camarillo(1)

1. Department of Bioengineering, Stanford University, Stanford, CA, 94305, USA.

2. Department of Biomedical Data Science, Stanford University, Stanford, CA, 94305, USA.

3. Department of Radiology, Stanford University, Stanford, CA, 94305, USA.

4. Department of Neurosurgery, Stanford University, Stanford, CA, 94305, USA.

Xianghao Zhan and Yiheng Li contributed equally to this work.

Corresponding author: Yuzhe Liu (yuzheliu@stanford.edu)

**Abstract**

Traumatic brain injury can be caused by various types of head impacts. However, due to different kinematic characteristics, many brain injury risk estimation models are not generalizable across the variety of impacts that humans may sustain. The current definitions of head impact subtypes are based on impact sources (e.g., football, traffic accident), which may not reflect the intrinsic kinematic similarities of impacts across the impact sources. To investigate the potential new definitions of impact subtypes based on kinematics, 3,161 head impacts from various sources including simulation, college football, mixed martial arts, and car racing were collected. We applied the K-means clustering to cluster the impacts on 16 standardized temporal features from head rotation kinematics. Then, we developed subtype-specific ridge regression models for cumulative strain damage (using the threshold of 15%), which significantly improved the estimation accuracy compared with the baseline method which mixed impacts from different sources and developed one model ($R^2$ from 0.7 to 0.9). To investigate the effect of kinematic features, we presented the top three critical features (maximum resultant angular acceleration, maximum angular acceleration along the z-axis, maximum linear acceleration along the y-axis) based on regression accuracy and used logistic regression to find the critical points for each feature that partitioned the subtypes. This study enables researchers to define head impact subtypes in a data-driven manner, which leads to more generalizable brain injury risk estimation.

**Introduction**

Traumatic brain injury (TBI) has become a growing public health problem with high mortality and morbidity, as well as a socio-economic problem causing enormous diagnosis and treatment expenses worldwide (James et al., 2019). In the United States, TBI contributed to a third of all injury-related deaths (Taylor et al., 2017), affecting 1.7 million people annually (Prins et al., 2012). Without directly causing death or disabilities, mild TBI (mTBI), which presents less severe symptoms and is harder to detect and diagnose, is also associated with severe consequences, leading to a form of silent pandemic: evidence has suggested that mTBI can lead to unconsciousness immediately after a head impact and can further result in post-concussive symptoms including cognitive deficits and emotional challenges (Shlosberg et al., 2010, Wallace and Morris, 2019) as well as increase the risk of long-term neurodegenerative diseases such as Parkinson's and Alzheimer's diseases (Doherty et al., 2016). The causes of TBI/mTBI are not limited to vehicular accidents and military combat, as they can also be sustained in contact sports such as football (Montenigro et al., 2017), mixed martial arts (MMA), ice hockey, water polo, lacrosse, and more (Caswell et al., 2017, Cecchi et al., 2019, Hernandez et al., 2015, O'Keeffe et al., 2020, Versace, 1971, Wilcox et al., 2014). Considering the severity and prevalence of TBI, fast diagnosis and early warning approaches (Beckwith et al., 2013) are crucial to preventing repetitive TBI, as intervention after early detection can attenuate injury severity to a significant extent (Ponsford et al, 2001).

Many previous studies have developed brain injury criteria (BIC) to estimate the risk of brain injury following head impacts. However, recent studies (Huber et al., 2021, Zhan et al., 2021) have found that different head impact subtypes (e.g., college football, MMA, car crashes) have different biomechanical characteristics and the generalizability of BIC may not be guaranteed across different head impact subtypes

(Zhan et al., 2021, Zhan et al., 2021b). To address this issue of generalizability, there are multiple options: first, we can better optimize BIC designs to enable more generalizable brain strain estimation models. For example, newly developed BIC such as rotational velocity change index (RVCI) (Yanaoka, et al., 2015) and the convolution of the impulse response for brain injury criterion (CIBIC) (Takahashi et al., 2017) have proven to be more generalizable across different sources of impacts (Zhan et al., 2021). Second, we can also address the generalizability issue by optimizing the datasets used to model BIC. For example, our previous study (Zhan et al., 2021c) has shown that by building a kinematics classifier to classify impacts from different sources and developing type-specific regression models, the accuracy of brain strain estimation can be significantly improved when compared with the strategy of combining various types of head impacts for one model. However, the classification strategy relies heavily on human definitions of head impact subtypes, which generally refer to the different impact sources and different sports. Although the study shows that head impacts originating from different sources generally manifest different spectral density distributions, not all impacts from the same source are alike in terms of features. Meanwhile, certain impacts in one subtype may bear more resemblance to another subtype. Therefore, it is worthwhile to investigate whether there are new head impact subtype definitions beyond the human definitions based on impact sources, and whether the new subtype definitions enable more accurate brain strain estimation.

To investigate potential new impact subtyping strategies, we used the kinematics of 3,161 head impacts originating from various sports/sources: head impact simulations, college football, MMA, and car racing crashes. We extracted both temporal and spectral features from the kinematics and applied the K-means clustering algorithm on 16 temporal features to learn the impact clusters in a fully data-driven manner. Then, we built the ridge regression model to estimate the cumulative strain damage (CSDM 15%, indicating the volume fraction of brain with MPS exceeding the threshold of 0.15 (Takhounts et al., 2008)) and compared the accuracy with 1) a baseline model developed with a mixture of different head impact subtypes and 2) a classification method which classifies the impacts based on impact sources and estimates CSDM with a type-specific ridge regression model. Finally, we analyzed the clustering results and found the critical features and critical points based on specific features to partition the impacts, which manifests the potential new definitions of head impact subtypes.

**Materials and Methods**

**1. Data description**

To investigate a variety of head impact subtypes, kinematics from a total of 3,161 head impacts from 4 different impact sources were collected: 2,130 lab-reconstructed impacts from a validated finite element head model of the Hybrid III anthropomorphic test dummy headform (dataset HM) (Giudice et al., 2019); 302 video-confirmed college football impacts recorded by the Stanford instrumented mouthguard (dataset CF) (Camarillo et al., 2013, Liu et al., 2020, Liu et al., 2021); 457 video-confirmed mixed-martial-arts impacts recorded by the Stanford instrumented mouthguard (dataset MMA) (O'Keeffe et al., 2020, Tiernan et al. 2020); and 272 reconstructed impacts from National Association for Stock Car Auto Racing (dataset NASCAR).

Finite-element (FE) modeling is a state-of-the-art tool for the biomechanical modeling of brain strain during head impacts, and we applied the validated KTH head model (Kleiven, 2007) to calculate CSDM. Different from maximal principal strain and strain rate type metrics that show the most severe deformation of brain tissue, CSDM describes the volume ratio of brain tissue that exerted brain strain exceeding a certain level and indicates severity of entire brain deformation. In this study, CSDM indicated the volume fraction of the brain with MPS exceeding the threshold of 0.15. Previous work has found CSDM to be an effective predictor of injury (Takhounts et al., 2008), especially for more severe head impacts (Sanchez et al, 2017), and has also been widely used to evaluate risk of head impacts (Fahlsted et al, 2021, Shi et al, 2020, Zhan et al., 2021, Zhan et al., 2021c).

**2. Kinematic feature extraction**

Previous studies have shown that the temporal features extracted from head kinematics, as well as multiple brain injury criteria (BIC) derived from the temporal information, are effective in brain strain estimation (Zhan et al., 2021, Zhan et al., 2021b). Further, the spectral densities have been found to be effective in classifying the source of head impacts (Zhan et al., 2021c). Therefore, in this study, we extracted both temporal and spectral features from the kinematics to leverage the advantages from both types of features.

The temporal features include the peaks of four types of kinematics: 1) the linear acceleration at the brain center of gravity $a(t)$, 2) the angular velocity $\omega(t)$, 3) the angular acceleration $\alpha(t)$ and 4) the angular jerk $j(t)$. For each kinematic type, the peaks of four channels were extracted: the components in three axes (X: posterior-to-anterior, Y: left-to-right, Z: superior-to-inferior) and the magnitude. To account for non-linearity in the relationship between kinematics and brain strain, we included the square-root and squared features based on the features previously mentioned, because our previous study has shown that the first-power, squared-root, and second-power features are the most predictive of brain strain (Zhan et al., 2021d).

The spectral features were extracted from these four kinematics types in a similar manner as shown previously (Zhan et al. 2021c), but we further optimized this feature extraction approach by adding the number of frequency windows. This allowed more detailed information in the low-frequency range to be extracted: we set 19 frequency windows with a width of 20Hz from 0Hz to 300Hz, and with another width of 50Hz from 300Hz to 500Hz (the Nyquist frequency). In each frequency window, the mean spectral density was extracted.

Furthermore, we extracted 15 BIC which can be regarded as the mathematically transformed temporal features because they were justified to be effective in brain strain estimation (Zhan et al., 2021): severity index (SI), head injury criterion (HIC), generalized acceleration model for brain injury threshold (GAMBIT), head impact power (HIP), principal component score (PCS), kinematic rotational brain injury criterion (BRIC), power rotation head injury criterion (PRHIC), Kleiven's linear combination (KLC), rotational injury criterion (RIC), brain injury criterion (BrIC), the combined probability of concussion (CP), rotational velocity change index (RVCI), the convolution of the impulse response for brain injury criterion (CIBIC), diffuse axonal multi-axis general evaluation (DAMAGE), and brain angle metric (BAM).

Therefore, a total of 367 features were extracted (48 temporal features: 4 kinematic types × 4 channels × 3 powers; 304 spectral features: 4 kinematic types × 4 channels × 19 frequency windows; 15 BIC) for the CSDM estimation.

## 3. Clustering model development

In this study, to find potential kinematic clusters beyond the current impact types defined by impact sources (e.g., college football, MMA, car crashes), we applied the K-means clustering algorithm to find clusters of kinematics in an entirely data-driven manner. K-means is an unsupervised clustering algorithm that minimizes the sum of the distance between each sample and its cluster centroid (MacQueen 1967). Without the information of the human-defined impact types, K-means learns the clusters completely based on the feature distribution. Given a set of n samples $(x_1, x_2, \ldots, x_n)$, the algorithm aims to partition the samples into K sets $\mathbf{S} = \{S_1, S_2, \ldots, S_K\}$, such that

$$argmin_S \sum_{i=1}^{K} \sum_{x \in S_i} ||x - \mu_i||^2$$

After randomly initializing the K centroids $(\mu_1, \mu_2, \ldots, \mu_K)$, the algorithm approaches this problem in an iterative manner until convergence:

1. Assign each sample to the cluster based on the nearest centroid:

$$S_i^{(t)} = \{x_p : ||x_p - \mu_i^{(t)}||^2 \leq ||x_p - \mu_j^{(t)}||^2, \forall j, 1 \leq j \leq K\}$$

2. Update the K centroids by calculating the cluster means:

$$\mu_i^{(t+1)} = \frac{1}{|S_i^{(t)}|} \sum_{x_j \in S_i^{(t)}} x_j$$

The clustering on all temporal and spectral features and on the 16 temporal features did not lead to significant CSDM regression accuracy difference (p>0.1). Therefore, we only performed the clustering on 16 temporal features (those temporal features excluding the 16 square-root features and the 16 squared features) which are defined in Table 1. Data standardization was performed before clustering to avoid the uneven weights of certain features due to the mismatch of the feature value ranges. We set the number of clusters K to be 2.

**Table 1. The 16 temporal features used in the K-means clustering.**

| No. | Feature meaning | Kinematics type |
|---|---|---|
| 1 | $\max\|a_x(t)\|$ | Linear acceleration |
| 2 | $\max\|a_y(t)\|$ | Linear acceleration |
| 3 | $\max\|a_z(t)\|$ | Linear acceleration |
| 4 | $\max\|a(t)\|$ | Linear acceleration |
| 5 | $\max\|\omega_x(t)\|$ | Angular velocity |
| 6 | $\max\|\omega_y(t)\|$ | Angular velocity |
| 7 | $\max\|\omega_z(t)\|$ | Angular velocity |
| 8 | $\max\|\omega(t)\|$ | Angular velocity |
| 9 | $\max\|\alpha_x(t)\|$ | Angular acceleration |
| 10 | $\max\|\alpha_y(t)\|$ | Angular acceleration |
| 11 | $\max\|\alpha_z(t)\|$ | Angular acceleration |
| 12 | $\max\|\alpha(t)\|$ | Angular acceleration |
| 13 | $\max\|j_x(t)\|$ | Angular jerk |
| 14 | $\max\|j_y(t)\|$ | Angular jerk |
| 15 | $\max\|j_z(t)\|$ | Angular jerk |
| 16 | $\max\|j(t)\|$ | Angular jerk |

**4. Regression model development**

For the regression of CSDM, to get better accuracy, we included all the 367 features as predictors which already included the non-linearity in the feature designs. We then built the ridge regression algorithm because it is an L2-regularized linear regression model, which balances the model complexity and generalizability, as well as the bias and the variance (Hoerl and Kennard, 1970), to enable a more generalizable CSDM regression model.

**5. Model assessment**

To evaluate the accuracy of the CSDM estimation across different head impact subtypes, two tasks were designed:

1) mixed-test: we mixed the head impacts originated from four different sources and partitioned the entire dataset into a training set with 80% data to train the clustering model and the regression models for each cluster, and a test set with 20% data to test the model accuracy;

2) leave-one-dataset-out: we regarded each of the three on-field datasets (CF/MMA/NASCAR) as unseen data and used the remaining two on-field datasets and the largest dataset (HM) to train the models and test on the unseen data. The reason why this task was performed is that researchers generally attach more significance to the performance with the on-field impacts and there are scenarios when models have to deal with impacts from unseen sources.

In the evaluation stage, the coefficients of determination ($R^2$) and the root mean squared error (RMSE) were calculated as the accuracy metrics. As the majority of the impacts come from dataset HM, directly calculating the metrics may lead to biased results, but as users of the metrics, we focus more on the performance on the on-field datasets. Therefore, we calculated the $R^2$ and RMSE for each type in the test data (HM/CF/MMA/NASCAR, based on the impact source) and took the averaged $R^2$ and RMSE as the ultimate accuracy metrics.

The two methods against which we compare the clustering method are 1) baseline method: train one single regression model on the 80% training data; 2) classification method: build a kinematics classifier to classify four head impact types based on impact sources, build the respective regression models for each head impact types based on the 80% training data, and use the type-specific regression models based on the kinematic type predictions in the testing phase. The classification model was shown effective in improving brain strain estimation accuracy in our previous study, and it can be regarded as a reference model based on the human definition of kinematic subtypes (Zhan et al., 2021c). In this study, instead of using the random forest classifier, we further optimized the classification algorithm performance by building a classifier on the larger set of temporal and spectral features with a multilayer perceptron (MLP).

For task 1, the experiments were done with 20 random dataset partitions to get robust results. A Wilcoxon signed-rank test was done on the $R^2$ results to test the statistical significance. The paired t-test was not used because the Shapiro Wilk test rejected the normal distribution assumption of some $R^2$ results. The hyperparameters of the models, such as the L2 penalty strength for each ridge regression model, were tuned based on five-fold cross-validation on the training data. The regression RMSE was the optimization goal.

**6. Effect of kinematic features on clustering**

In the previous sections, we introduced the clustering on 16 temporal features as there was no significant difference on regression with the clustering on all features. To investigate the effect of particular kinematic features, we further analyzed the clustering results based on each of the 16 temporal features. We selected the top individual features that partition the impacts to two clusters, which as a result, led to the highest CSDM regression accuracy. The top three features were selected based on the mean CSDM regression accuracy among 20 experiments. Then, further analysis was done on finding critical points on these three critical features that partition the impacts into two clusters based on the entire dataset with 3,161 impacts. Because clustering with individual features is equivalent to partitioning the impacts into two with critical points, we compute the critical points corresponding to the top individual features. To ensure robust results, for each critical feature, we first got the robust clustering labels by performing K-means clustering 100 times and labeled each impact based on the mode of the 100 clustering results. Then, we fitted a logistic regression (LR) model and found the critical point c where the probability of the feature value c being in one cluster equals 0.5:

$$P(Y = 1|X = c) = P(Y = 2|X = c) = 0.5$$

**Results**

In the mixed-test task (task 1), we tested the performance of the K-means clustering method (labeled as KMeans), which is shown in Fig. 1. The CSDM regression method with the K-means clustering outperformed the baseline method (mix all training impacts and train one model) and the classification method, with statistical significance (p<0.001) based on the averaged $R^2$ on four types of head impacts (HM/CF/MMA/NASCAR). Further, the K-means clustering method led to significant accuracy improvement on all of the four impact sources when viewed individually (p<0.001).

Then, we picked up the top three critical features that led to the highest mean CSDM regression accuracy in 20 experiments, which are shown in Fig. 2. With the K-means clustering on these three critical features, the regression accuracy was also generally improved, although the clustering on the second and third critical features did not show statistically significant improvement over the classification models (p>0.1). Furthermore, in Fig. 1, we showed the model performances on the specific types of on-field head impacts (CF/MMA/NASCAR) because the ultimate goal of the estimation model is for the on-field impacts. The results and corresponding statistical tests manifested that clustering on the 16 temporal features or on the first critical feature can lead to statistically significant accuracy improvement in CSDM estimation for all three types of on-field data (CF/MMA/NASCAR, p<0.001), as well as for the averaged results over four types of data (p<0.001), while clustering on the second and third critical features led to non-decreasing CSDM estimation accuracy when compared with the baseline method and the classification method. It should be noted that clustering on 16 temporal features or on the first critical feature did not lead to a statistically significant difference in the accuracy of the averaged results over four types of impacts (p>0.05).

The three critical features we found are the maximum resultant angular acceleration, the maximum angular acceleration along the z-axis, and the maximum linear acceleration along the y-axis. On each of these three critical features, we fitted the logistic regression model for the two clusters and reported the critical point values in Table 1. To better visualize the brain strain profile of all the 4,124 brain elements models by the KTH model, we also plotted the heatmaps of all the 4,124 impacts with the clusters labeled in Fig. 3. It is shown that the critical points for each critical feature partitioned the impacts into two clusters with generally different levels of CSDM but the subtype partitions based on the clustering on the critical features were not strictly associated with the high/low CSDM values. It is important to note that the clusters separated by the critical points were the same as the robust K-means clustering results over 100 experiments because the clustering was performed on one dimension based on Euclidean distance. Furthermore, on task 2 (leave-one-dataset-out), the CSDM regression accuracy in terms of RMSE is shown in Fig. 4. The results indicate that although the prediction task is quite challenging when we completely leave one type of head impacts out, with the K-means clustering the CSDM regression accuracy can be improved. The strategy of finding impact subtypes with clustering may facilitate the risk estimation when there is a missing dataset of a source of impacts on which we want to estimate the brain injury risk.

**Discussion**

In this study, we applied the K-means clustering to find potential subtypes among impacts from four different sources/sports in a completely data-driven manner. The K-means clustering algorithms found the two clusters based on the Euclidean distance on 16 standardized temporal features. With these subtypes found by clustering, we performed CSDM estimation. The results manifested that the clustering and subtype-specific CSDM regression can lead to a more accurate estimation of CSDM, when compared with 1) the conventional strategy of using all data to train one CSDM regression model and 2) our previously proposed classification strategy to build type-specific regression models based on the impact sources (Zhan et al. 2021c). Compared with the classification method, the novel clustering method proposed in this study can be applied to better address the generalizability issue of the brain injury risk estimation model, which has been proposed by Zhan et al. (Zhan et al. 2021). Furthermore, the critical features and the corresponding critical point values related to the subtype separation were investigated, which turned out to improve the CSDM estimation accuracy. This completely data-driven clustering enables researchers to develop new definitions of impact subtypes that go beyond the current human definitions of impact types based on sports/sources, as not all impacts from the same source are alike. This finding indicates that there may be different modes of injury mechanisms for impacts above or below the critical points.

According to the results, in the CSDM-based TBI risk estimation, both 1) one general linear regression model and 2) the type-specific linear regression method with four type-specific linear regression models based on impact source classification (Zhan et al. 2021c) are not as accurate as the two-segment linear models built for impacts in the two ranges partitioned by the critical features and critical points. The K-means clustering method, as well as the classification method, are done completely based on the kinematics features without knowing the brain strain (CSDM). Therefore, we clustered on the predictors (kinematics), but not the relationship between the predictors and the response (CSDM). Without developing the two-segment linear models based on the predictor-response relationship, the clustering method found the critical points and critical features to partition the potential subtypes. Usually, replacing a linear model with a two-segment linear model improves the fit on the training impacts, but may not improve the regression accuracy on unseen test data, because although higher degrees of freedom lead to better fit, this does not necessarily lead to better model predictability (Friedman et al., 2001). In this study, what is remarkable is that after performing subtyping with clustering, the CSDM regression accuracy tested on the unseen test impacts was significantly improved. This observation indicates that there can be significantly different injury mechanisms for impacts partitioned by the critical features and critical points.

As a viscoelastic material, the response of brain strain is highly influenced by the traces of head kinematics (Zhao and Ji, 2017). Therefore, the relationship between the kinematics features and brain strain may vary along with different characteristics of traces. In this study, the better prediction of CSDM using K-means clustering suggests the potential of different mechanisms that govern how the inertial loading leads to the CSDM. Besides, in the previous studies (Carlsen et al., 2021, Gabler et al., 2018), the duration of impulse has been found to subtype the relationship between brain strain peak and kinematics peak (95th percentile maximum principal strain (MPS95) correlates with angular velocity peak when the duration is short, with angular acceleration peak when duration is long). However, it is unclear if the duration of the impulse results in the subtyping is based on the clusters because of the difference between CSDM and MPS95.

In this study, we have reported the regression results of CSDM based on the baseline, classification, and clustering methods. In our preliminary experiments, we have also tested the effect of kinematics clustering on the regression of the MPS95, which is also regarded as an important strain-based injury predictor. However, the kinematics clustering did not show a statistically significant difference in MPS95 estimation accuracy. Physiologically, MPS and CSDM describe brain strain in different ways: generally, MPS95 represents the highest strain across the entire brain and CSDM represents the overall percentage of high strain regions in the brain (Takhounts, et al., 2008). Therefore, the relationship between kinematics and MPS95/CSDM can be different. As we probe the relationship with ridge regression, different feedback has been observed: there may be varying linear relationships for CSDM but this relationship may not exist for MPS95. These observations warrant further analysis among researchers to further investigate the mechanism of causing high strain.

In this study, we reported the cases when we consider using the K-means to partition all impacts into two clusters. We have also considered the three-cluster case in our preliminary experiments with all the temporal and spectral features, which turned out to show similar significant accuracy improvement over the baseline method and the classification method but slightly inferior CSDM estimation accuracy when compared to the two-cluster case. Therefore, we focused on the binary divisions of impacts with KMeans clustering.

Although this study shows that it is possible to develop new definitions of impact subtypes that enable better CSDM estimation, there are several limitations. First, we only performed the clustering on the 16 temporal features to make the new subtype definitions easy to interpret and readily understandable and applicable for researchers without any further requirement of kinematics transforms or computation. It is possible that clustering on a wider range of features (e.g., spectral densities) may lead to even better results at the sacrifice of interpretability. Second, although we have verified that the subtypes found by KMeans clustering contribute to better CSDM estimation, how exactly the different subtypes differ in injury mechanism is still not fully understood. Imaging results can be incorporated to further investigate the subtypes partitioned by the kinematics critical features and critical points. Furthermore, we used the KTH model to calculate brain strain. This model is limited when compared to recently developed state-of-the-art finite element head models (FEHM) (Li et al. 2021). For example, the KTH model does not take the gyri or sulci into modeling, which have been shown to have significant influences on FEHM behavior. In the future, more recently developed FEHMs can be applied to validate the results on brain strain.

**Conclusion**

This study applies the K-means clustering to cluster the impacts originated from various sources and sports into subtypes in a data-driven manner. The clustering and subtype-specific modeling significantly improve the regression accuracy for cumulative strain damage when compared with the practice of mixing different types of head impacts. Based on the clustering and logistic regression, critical features and critical points were found. The results may give potential new definitions of head impact subtypes related to varying mechanisms.

**Acknowledgement**

This research was supported by the Pac-12 Conference's Student-Athlete Health and Well-Being Initiative, the National Institutes of Health (R24NS098518), Taube Stanford Children's Concussion Initiative and Stanford Department of Bioengineering.

**Conflict of interest statement**

The authors declare no commercial or financial relationships related to this study as a potential conflict of interest.

**Reference**

1. Beckwith, J.G., Greenwald, R.M., Chu, J.J., Crisco, J.J., Rowson, S., Duma, S.M., Broglio, S.P., McAllister, T.W., Guskiewicz, K.M., Mihalik, J.P. and Anderson, S., 2013. Timing of concussion diagnosis is related to head impact exposure prior to injury. Medicine and science in sports and exercise, 45(4), p.747.

2. Camarillo, D.B., Shull, P.B., Mattson, J., Shultz, R. and Garza, D., 2013. An instrumented mouthguard for measuring linear and angular head impact kinematics in American football. Annals of biomedical engineering, 41(9), pp.1939-1949.

3. Carlsen, R.W., Fawzi, A.L., Wan, Y., Kesari, H. and Franck, C., 2021. A quantitative relationship between rotational head kinematics and brain tissue strain from a 2-D parametric finite element analysis. Brain Multiphysics, 2, p.100024.

4. Caswell, S.V., Lincoln, A.E., Stone, H., Kelshaw, P., Putukian, M., Hepburn, L., Higgins, M. and Cortes, N., 2017. Characterizing verified head impacts in high school girls' lacrosse. The American journal of sports medicine, 45(14), pp.3374-3381.

5. Cecchi, N.J., Monroe, D.C., Fote, G.M., Small, S.L. and Hicks, J.W., 2019. Head impacts sustained by male collegiate water polo athletes. PloS one, 14(5), p.e0216369.

6. Doherty, C.P., O'Keefe, E., Wallace, E., Loftus, T., Keaney, J., Kealy, J., Humphries, M.M., Molloy, M.G., Meaney, J.F., Farrell, M. and Campbell, M., 2016. Blood–brain barrier dysfunction as a hallmark pathology in chronic traumatic encephalopathy. Journal of Neuropathology & Experimental Neurology, 75(7), pp.656-662.

7. Fahlstedt, M., Abayazid, F., Panzer, M.B., Trotta, A., Zhao, W., Ghajari, M., Gilchrist, M.D., Ji, S., Kleiven, S., Li, X. and Annaidh, A.N., 2021. Ranking and rating bicycle helmet safety performance in oblique impacts using eight different brain injury models. Annals of biomedical engineering, 49(3), pp.1097-1109.

8. Friedman, J., Hastie, T. and Tibshirani, R., 2001. The elements of statistical learning (Vol. 1, No. 10). New York: Springer series in statistics. pp. 219-227.

9. Gabler, L.F., Joodaki, H., Crandall, J.R. and Panzer, M.B., 2018. Development of a single-degree-of-freedom mechanical model for predicting strain-based brain injury responses. Journal of biomechanical engineering, 140(3), p.031002.

10. Giudice, J.S., Park, G., Kong, K., Bailey, A., Kent, R. and Panzer, M.B., 2019. Development of open-source dummy and impactor models for the assessment of American football helmet finite element models. Annals of biomedical engineering, 47(2), pp.464-474.

11. Hernandez, F., Wu, L.C., Yip, M.C., Laksari, K., Hoffman, A.R., Lopez, J.R., Grant, G.A., Kleiven, S. and Camarillo, D.B., 2015. Six degree-of-freedom measurements of human mild traumatic brain injury. Annals of biomedical engineering, 43(8), pp.1918-1934.

12. Hoerl, A.E. and Kennard, R.W., 1970. Ridge regression: Biased estimation for nonorthogonal problems. Technometrics, 12(1), pp.55-67.

13. James, S.L., Theadom, A., Ellenbogen, R.G., Bannick, M.S., Montjoy-Venning, W., Lucchesi, L.R., Abbasi, N., Abdulkader, R., Abraha, H.N., Adsuar, J.C. and Afarideh, M., 2019. Global, regional, and national burden of traumatic brain injury and spinal cord injury, 1990–2016: a systematic analysis for the Global Burden of Disease Study 2016. The Lancet Neurology, 18(1), pp.56-87.

14. Kleiven, S., 2007. Predictors for traumatic brain injuries evaluated through accident reconstructions. Stapp car crash J, 51(81), pp.81-114.

15. Li, X., Zhou, Z. and Kleiven, S., 2021. An anatomically detailed and personalizable head injury model: Significance of brain and white matter tract morphological variability on strain. Biomechanics and modeling in mechanobiology, 20(2), pp.403-431.

16. Liu, Y., Domel, A.G., Yousefsani, S.A., Kondic, J., Grant, G., Zeineh, M. and Camarillo, D.B., 2020. Validation and comparison of instrumented mouthguards for measuring head kinematics and assessing brain deformation in football impacts. Annals of Biomedical Engineering, 48(11), pp.2580-2598.

17. Liu, Y., Zhan, X., Domel, A.G., Fanton, M., Zhou, Z., Raymond, S.J., Alizadeh, H.V., Cecchi, N.J., Zeineh, M. and Grant, G., 2020. Theoretical and numerical analysis for angular acceleration being determinant of brain strain in mTBI. arXiv preprint arXiv:2012.13507.

18. Liu, Y., Domel, A.G., Cecchi, N.J., Rice, E., Callan, A.A., Raymond, S.J., Zhou, Z., Zhan, X., Zeineh, M., Grant, G. and Camarillo, D.B., 2021. Time window of head impact kinematics measurement for calculation of brain strain and strain rate in American football. Annals of Biomedical Engineering. Accepted.

19. MacQueen, J., 1967, June. Some methods for classification and analysis of multivariate observations. In Proceedings of the fifth Berkeley symposium on mathematical statistics and probability (Vol. 1, No. 14, pp. 281-297).

20. Montenigro, P.H., Alosco, M.L., Martin, B.M., Daneshvar, D.H., Mez, J., Chaisson, C.E., Nowinski, C.J., Au, R., McKee, A.C., Cantu, R.C. and McClean, M.D., 2017. Cumulative head impact exposure predicts later-life depression, apathy, executive dysfunction, and cognitive impairment in former high school and college football players. Journal of neurotrauma, 34(2), pp.328-340.

21. O'Keeffe, E., Kelly, E., Liu, Y., Giordano, C., Wallace, E., Hynes, M., Tiernan, S., Meagher, A., Greene, C., Hughes, S. and Burke, T., 2020. Dynamic blood–brain barrier regulation in mild traumatic brain injury. Journal of neurotrauma, 37(2), pp.347-356.

22. Prins, M.L. and Giza, C.C., 2012. Repeat traumatic brain injury in the developing brain. International journal of developmental neuroscience, 30(3), pp.185-190.

23. Ponsford, J., Willmott, C., Rothwell, A., Cameron, P., Ayton, G., Nelms, R., Curran, C. and Ng, K., 2001. Impact of early intervention on outcome after mild traumatic brain injury in children. Pediatrics, 108(6), pp.1297-1303.

24. Sanchez, E.J., Gabler, L.F., McGhee, J.S., Olszko, A.V., Chancey, V.C., Crandall, J.R. and Panzer, M.B., 2017. Evaluation of head and brain injury risk functions using sub-injurious human volunteer data. Journal of neurotrauma, 34(16), pp.2410-2424.

25. Shi, L., Han, Y., Huang, H., Davidsson, J. and Thomson, R., 2020. Evaluation of injury thresholds for predicting severe head injuries in vulnerable road users resulting from ground impact via detailed accident reconstructions. Biomechanics and modeling in mechanobiology, 19(5), pp.1845-1863.

26. Shlosberg, D., Benifla, M., Kaufer, D. and Friedman, A., 2010. Blood–brain barrier breakdown as a therapeutic target in traumatic brain injury. Nature Reviews Neurology, 6(7), pp.393-403.

27. Takahashi, Y. and Yanaoka, T., 2017. A study of injury criteria for brain injuries in traffic accidents. In 25th ESV conference, Paper (No. 17-0040).

28. Takhounts, E.G., Ridella, S.A., Hasija, V., Tannous, R.E., Campbell, J.Q., Malone, D., Danelson, K., Stitzel, J., Rowson, S. and Duma, S., 2008. Investigation of traumatic brain injuries using the next generation of simulated injury monitor (SIMon) finite element head model (No. 2008-22-0001). SAE Technical Paper.

29. Taylor, C.A., Bell, J.M., Breiding, M.J. and Xu, L., 2017. Traumatic brain injury–related emergency department visits, hospitalizations, and deaths—United States, 2007 and 2013. MMWR Surveillance Summaries, 66(9), p.1.

30. Tiernan, S., Meagher, A., O'Sullivan, D., O'Keeffe, E., Kelly, E., Wallace, E., Doherty, C.P., Campbell, M., Liu, Y. and Domel, A.G., 2020. Concussion and the severity of head impacts in mixed martial arts. Proceedings of the Institution of Mechanical Engineers, Part H: Journal of engineering in medicine, 234(12), pp.1472-1483.

31. Versace, J., 1971. A review of the severity index.

32. Wallace, T. and Morris, J., 2019. Development and Testing of a Technology Enhanced Intervention to Support Emotion Regulation in Military mTBI with PTSD. Archives of Physical Medicine and Rehabilitation, 100(7), p.e5.

33. Wilcox, B.J., Machan, J.T., Beckwith, J.G., Greenwald, R.M., Burmeister, E. and Crisco, J.J., 2014. Head-impact mechanisms in men's and women's collegiate ice hockey. Journal of athletic training, 49(4), pp.514-520.

34. Yanaoka, T., Dokko, Y. and Takahashi, Y., 2015. Investigation on an injury criterion related to traumatic brain injury primarily induced by head rotation (No. 2015-01-1439). SAE Technical Paper.

35. Zhao, W. and Ji, S., 2017. Brain strain uncertainty due to shape variation in and simplification of head angular velocity profiles. Biomechanics and modeling in mechanobiology, 16(2), pp.449-461.

36. Zhan, X., Li, Y., Liu, Y., Domel, A.G., Alizadeh, H.V., Raymond, S.J., Ruan, J., Barbat, S., Tiernan, S., Gevaert, O. and Zeineh, M.M., 2021. The relationship between brain injury criteria and brain strain across different types of head impacts can be different. Journal of the Royal Society Interface, 18(179), p.20210260.

37. Zhan, X., Liu, Y., Raymond, S.J., Alizadeh, H.V., Domel, A., Gevaert, O., Zeineh, M., Grant, G. and Camarillo, D.B., 2021. Rapid Estimation of Entire Brain Strain Using Deep Learning Models. IEEE Transactions on Biomedical Engineering.

38. Zhan, X., Li, Y., Liu, Y., Cecchi, N.J., Raymond, S.J., Zhou, Z., Alizadeh, H.V., Ruan, J., Barbat, S., Tiernan, S. and Gevaert, O., 2021. Classification of head impacts based on the spectral density of measurable kinematics. arXiv preprint arXiv:2104.09082.

39. Zhan, X., Li, Y., Liu, Y., Domel, A.G., Alidazeh, H.V., Zhou, Z., Cecchi, N.J., Tiernan, S., Ruan, J., Barbat, S. and Gevaert, O., 2021. Predictive Factors of Kinematics in Traumatic Brain Injury from Head Impacts Based on Statistical Interpretation. Annals of Biomedical Engineering. Accepted.

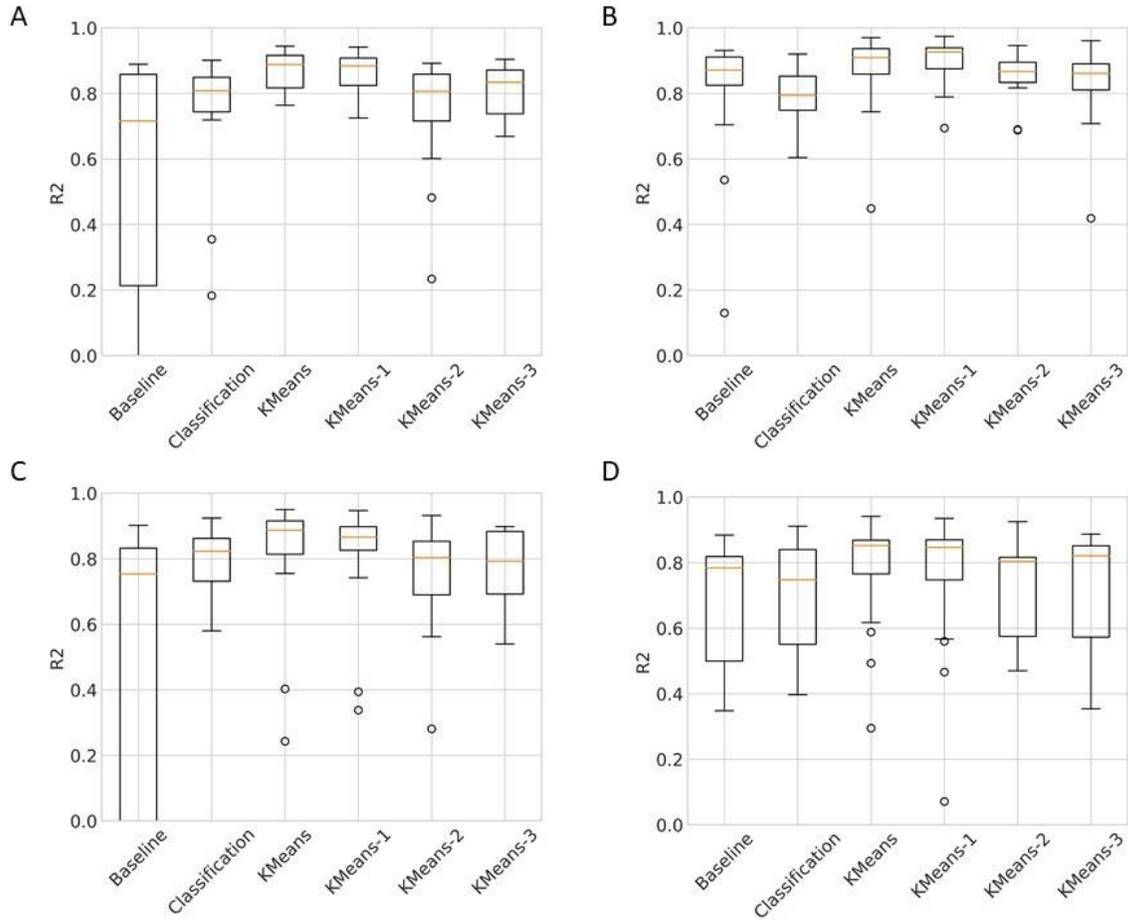

**Figure 1. The CSDM regression $R^2$ with ridge regression and different methods on task 1.** (A) The averaged CSDM regression $R^2$ over four datasets (HM/CF/MMA/NASCAR). The CSDM regression $R^2$ on the on-field test impacts originated from dataset CF (B), dataset MMA (C), and dataset NASCAR (D). KMeans-1/2/3: The K-means clustering method with the first/second/third critical feature: the maximum resultant angular acceleration, the maximum angular acceleration along the z-axis, the maximum linear acceleration along the y-axis.

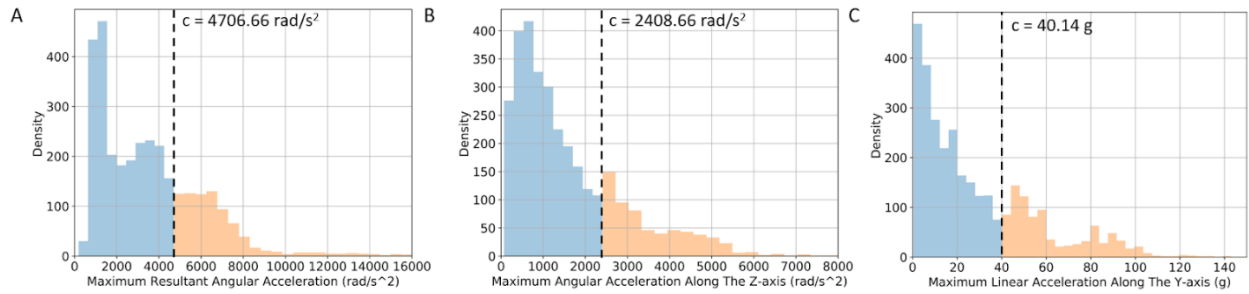

**Figure 2. The distribution of the two clusters partitioned by the critical points for the critical features.** The critical points c for the critical features: the maximum resultant angular acceleration (A), the maximum angular acceleration along the z-axis (B), the maximum linear acceleration along the y-axis (C)

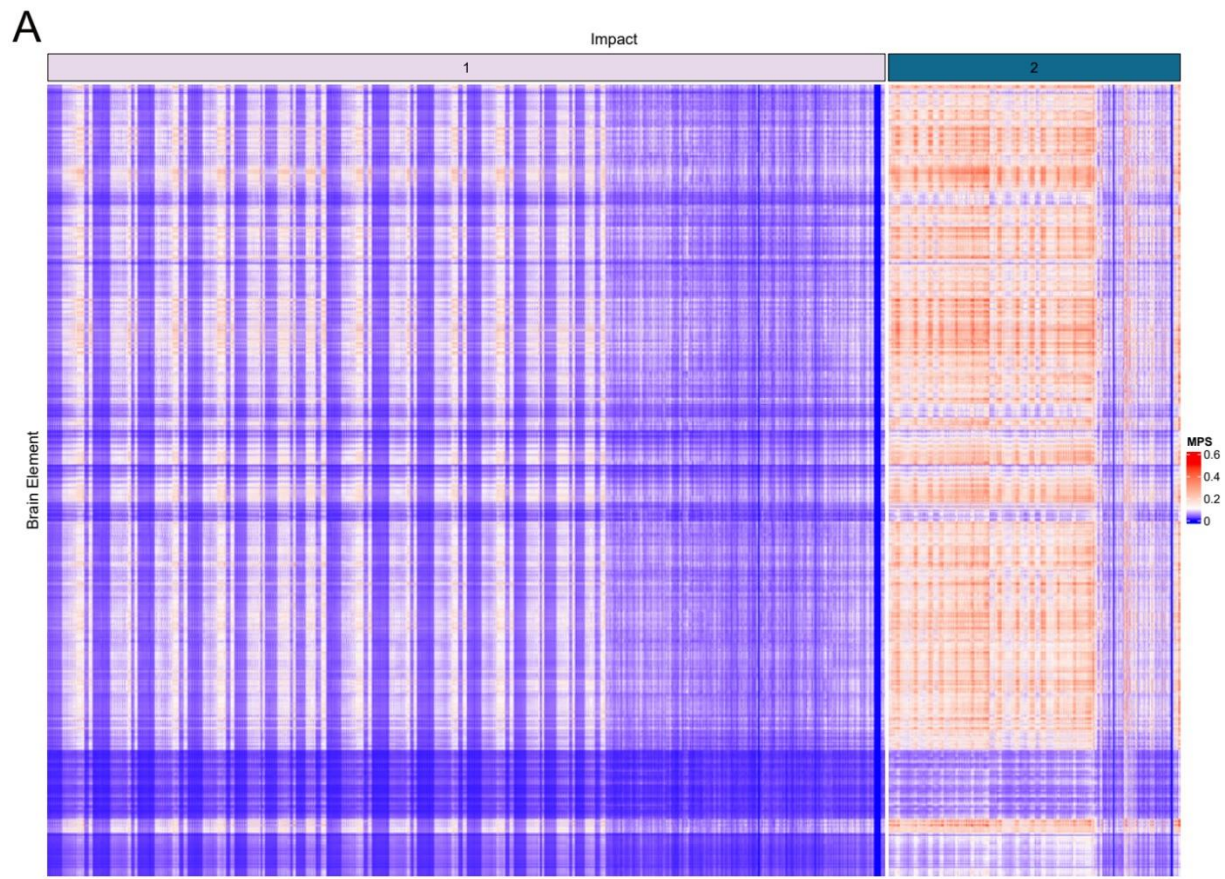

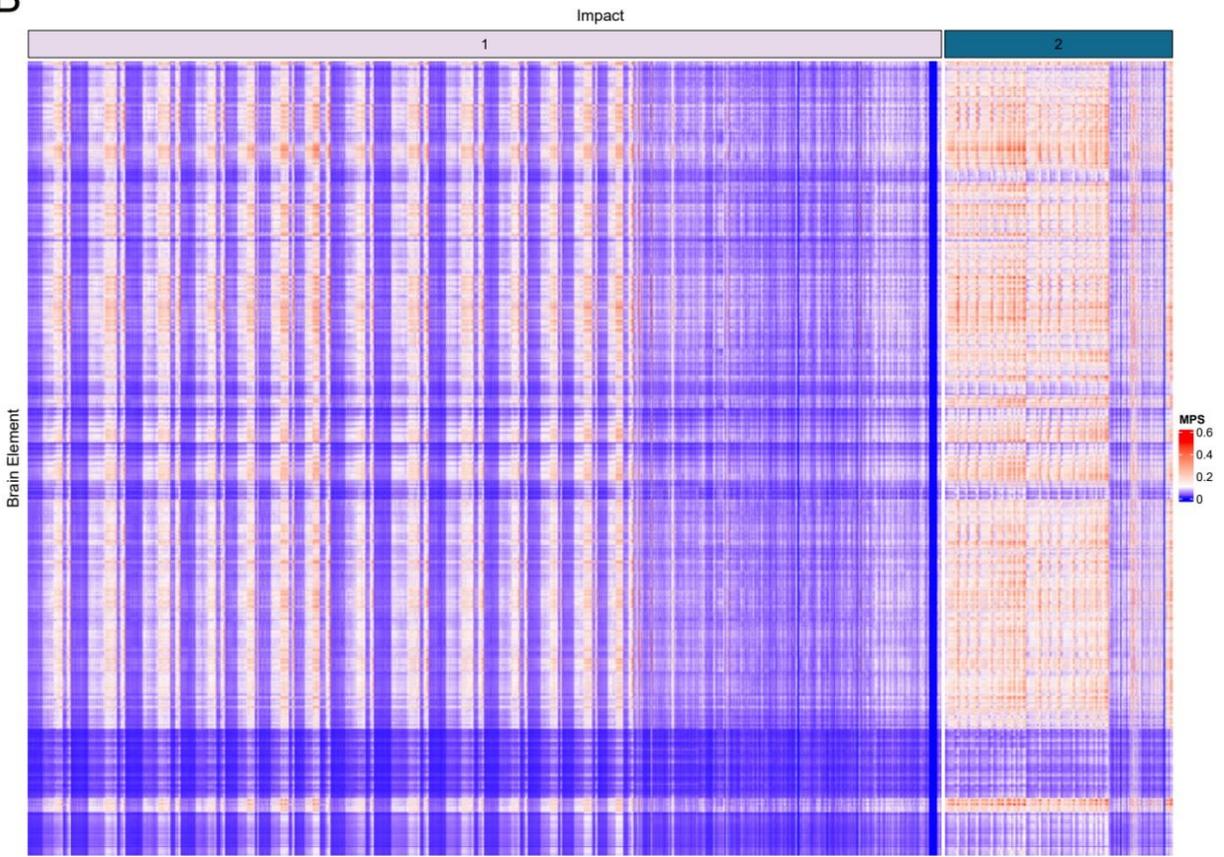

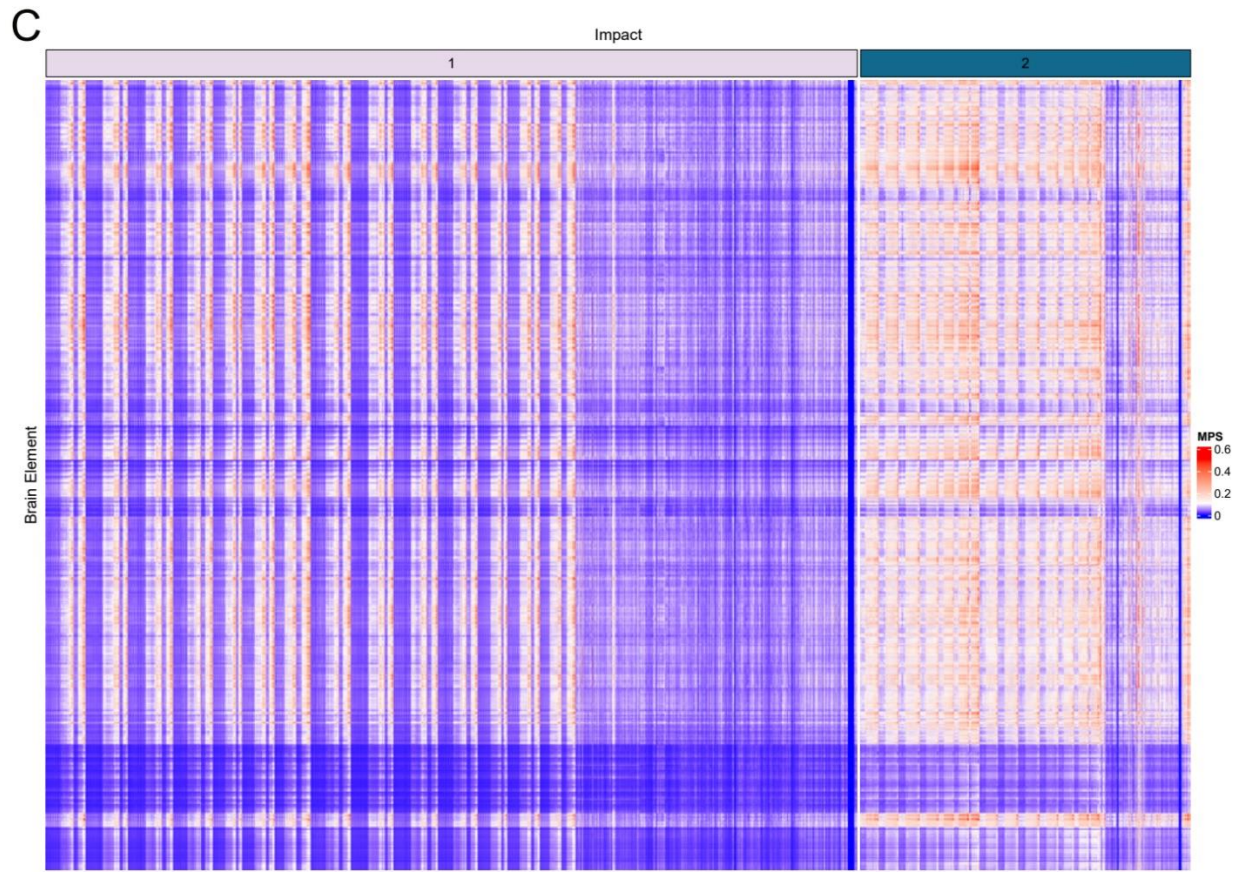

**Figure 3. The heatmap of the brain strain calculated by the KTH model and the assignment of clusters based on each of the three critical features.** The heatmap of the brain strain with the clustering on the maximum resultant angular acceleration (A), the maximum angular acceleration along the z-axis (B), the maximum linear acceleration along the y-axis (C).

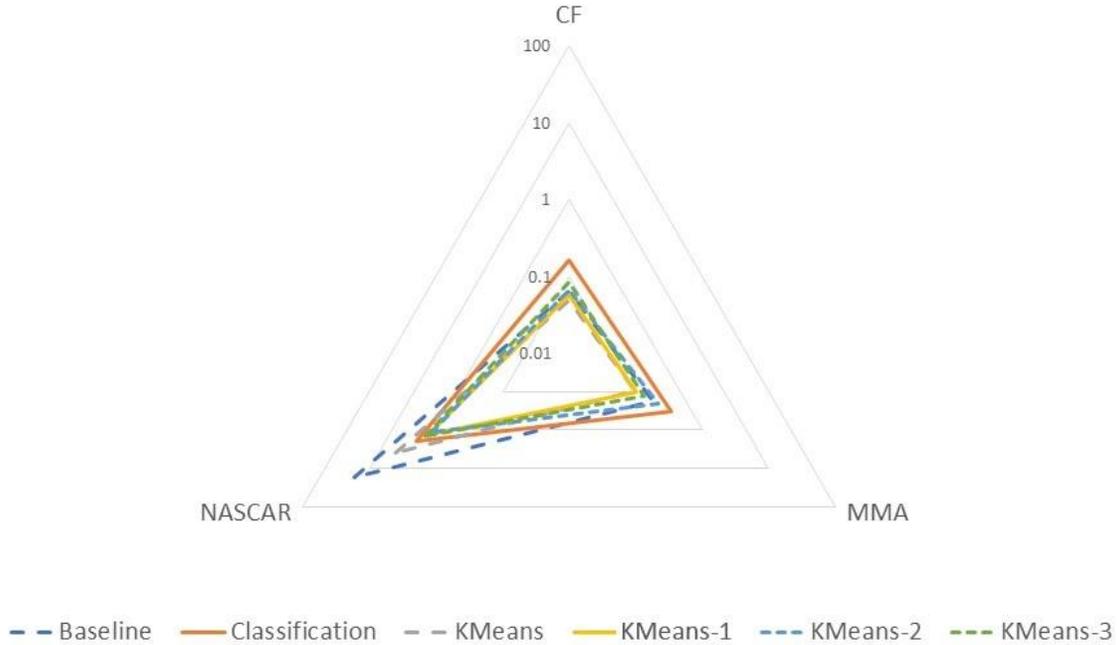

| Test dataset | Baseline | Classification | KMeans | KMeans-1 | KMeans-2 | KMeans-3 |
|---|---|---|---|---|---|---|
| CF | 0.063 | 0.160 | **0.049** | 0.057 | 0.066 | 0.085 |
| MMA | 0.183 | 0.340 | **0.097** | 0.103 | 0.214 | 0.134 |
| NASCAR | 16.856 | 1.957 | 4.084 | 1.328 | **1.132** | 1.429 |

**Figure 4. The CSDM regression accuracy on task 2.** Each of the three on-field datasets (CF/MMA/NASCAR) is once used as the test dataset while the remaining datasets are combined with the dataset HM to train the regression models. KMeans-1/2/3: The K-means clustering method with the first/second/third critical feature: the maximum resultant angular acceleration, the maximum angular acceleration along the z-axis, the maximum linear acceleration along the y-axis.